\documentclass[pra,showpacs,twocolumn,showkeys]{revtex4-1}
\usepackage[utf8]{inputenc}

\usepackage{graphicx}
\usepackage{amsmath}
\usepackage{amsfonts}
\usepackage{amssymb}
\usepackage{physics}
\usepackage{setspace}

\usepackage{graphicx}
\graphicspath{ {./images/}}

\usepackage[T1]{fontenc}
\usepackage[font=small,labelfont=bf,tableposition=top]{caption}

\DeclareCaptionLabelFormat{andtable}{#1~#2  \&  \tablename~\thetable}

\begin{document}
\newcommand{\bstfile}{osa}
\newcommand{\bibs}{d:/Dropbox/Dad/Mark/References/BibFile}
\title{Collecting data with a Mobile Phone: studies of mechanical laws such as energy and momentum conservation}
\author{Maximillian Hart}
\email{maximillian.hart@wsu.edu}
\author{Mark G. Kuzyk}
\email{kuz@wsu.edu}
\affiliation{Department of Physics and Astronomy, Washington State University, Pullman, Washington  99164-2814 \\ \today}
\begin{abstract}
We use videos taken with a mobile phone to study conservation of energy, conservation of momentum and the work-energy theorem by analyzing the collision of a cue ball and the eight ball.  A video of the full time sequence, starting from before the cue ball is struck until well after the collision, is recorded with a mobile phone.  The video is imported into Origin (free to teachers and students taking a class),\cite{origin2020} where the coordinates of the balls are digitized frame-by-frame using the free Video Extractor app.  From this data, the velocities are determined as a function of time, and used to determine the energy and momentum of each ball.  The data suggests that the balls slip through part of their motion when rotating at an angular velocity different than the rolling angular velocity, so angular momentum and frictional torque must be considered.  Such experiments require no specialized equipment other than a device to take a video, and the act of digitizing the data provides the student with visual reinforcement of the physics.  Experimentation by students outside of campus can be fun for them and provides a useful alternative to classes when in-lab experiments are not practical.  Use of a mobile phone in this way is generally applicable to any other mechanical phenomena that involves motion.
\end{abstract}

\maketitle

\tableofcontents

\section{Introduction}

With the almost-universal availability of smart phones, it is possible for students to conduct experiments at home, which in the past required stationary laboratories with expensive equipment.  One useful app is Phyphox, which uses a smart phone's built-in sensors to gather data.\cite{stack20.01}  For example, the phone can be thrown to determine the equations of motion of the trajectory from Phyphox's accelerometer.


Castro-Palacio and coworkers used the accelerometer in a cell phone to study oscillations of a mobile phone attached to an air track.\cite{Palacio13.01}  Another clever experiment used a mouse ball to collect data,\cite{ochoa97.01} which was later extended by Ng and Ang.\cite{ng05.01}  Other references to the literature where researchers adapted consumer hardware to physics experiments was pointed out by Kuhn.\cite{kuhn14.01}  Such experiments require that the cell phone be built into the experiment.

An alternative way to collect data is to record a video of the motion under study, then digitize the data for analysis.  We use the Video Extractor App from OriginLab Corporation,\cite{origin2020,origin20.10} which digitizes the data frame-by-frame.  This is similar in spirit to the work of Monsoriu and coworkers, who automated the analysis of videos with image recognition\cite{Palacio13.01} and that of Shamim et al.\cite{shami10.10}  Matavan and coworkers studied collisions of snooker balls with a high-speed camera and image recognition software, and like our work, studied slipping and rolling friction, but did not consider the work-energy theorem nor linear momentum conservation.\cite{matha09.01}  In contrast, our focus is geared to doing experiments at home.

Origin and Video Extractor are available for free from OriginLab Corporation\cite{origin20.10} for the students and instructors provided that the software is used for a course.  A free alternative is Tracker.\cite{tracker20.01}  Vernier is a commercial product that retails for \$49.\cite{vernier20.01}  We choose to use Video Extractor with Origin because it includes sophisticated data analysis packages, IMSL libraries, built-in C and Python, and creates professional-quality graphics.  The alternatives require that data be exported for more sophisticated analysis.

This paper resulted from a junior-level classical mechanics class that used Origin for student projects.  While digitizing data by mouse click is more time consuming than image recognition algorithms, it requires the student to be in direct contact with the data, which builds a more intimate understanding of data analysis and error analysis.  Finally, it is far simpler to take a video of an event than to attach a cell phone to the event.


\section{Experiment}

\begin{figure}[h]
    \centering
       \includegraphics[width=3.35 in]{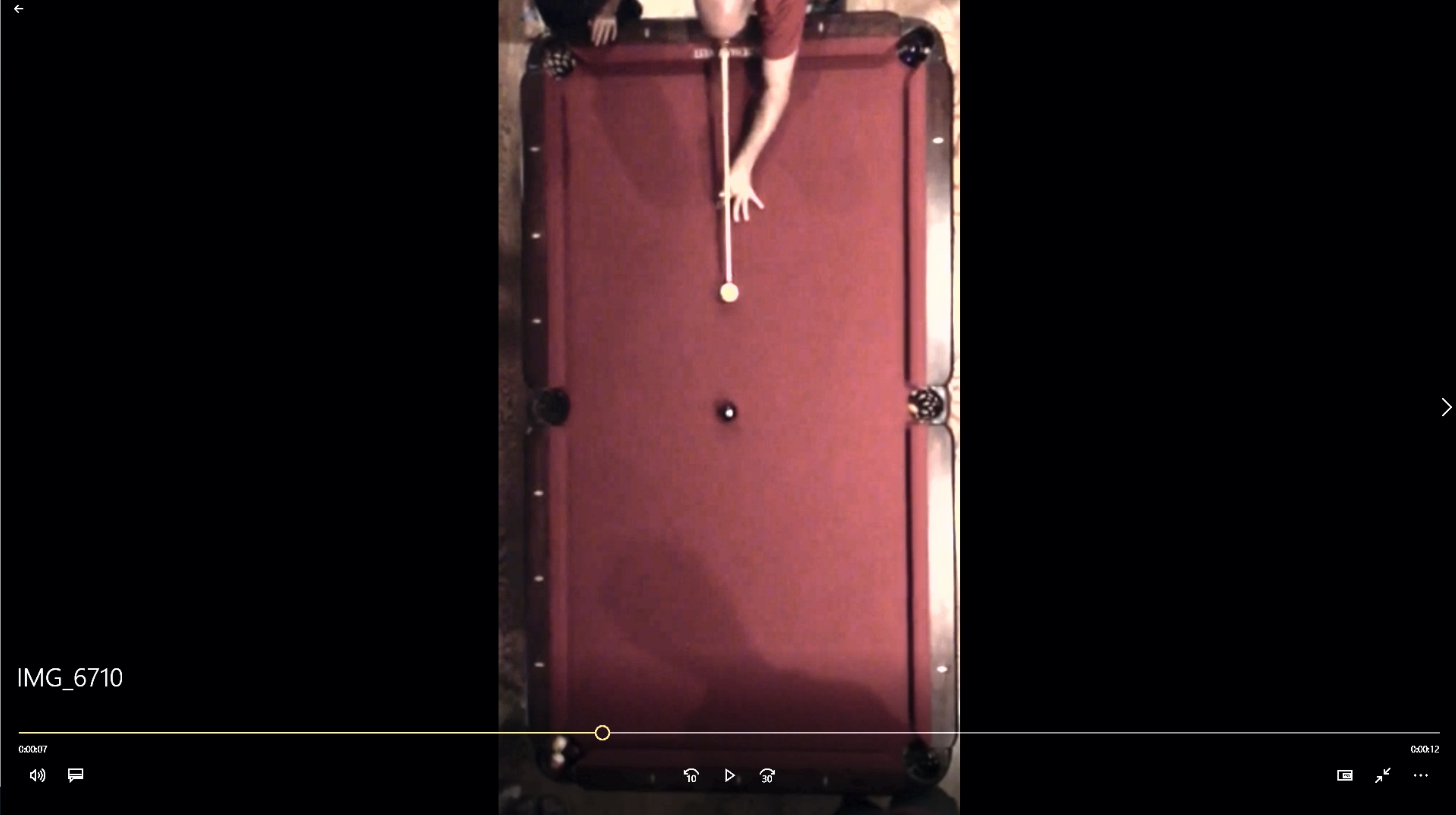}
    \caption{One frame of the video used to study the collisions of billiard balls.}
    \label{ref:Frame}
\end{figure}

In the present paper, we use Video Extractor with Origin to study energy and momentum conservation in collisions, rolling and sliding friction, angular momentum in the presence of slipping torque, and dissipative forces.  Our work is similar to past work of Williamson\cite{willi00.01} that used videos with a strobe to map out the trajectory, but our experiments do not require a strobe or an air table.  Such experiments help students visualize the process, which enhances learning.\cite{escal97.01} Our experiment is an efficient learning tool because one simple video can be used to study multiple principles and leads to several lab reports.

The motion of the cue ball before and after the collision with the eight ball was recorded with an iPhone 6 set to a frame rate of 240 fps.  As a check we took a video of a clock with a second hand, and found that the actual frame rate was 120 fps, which we use for the analysis.  This observation teaches the important lesson that the experimenter cannot always trust an instrument.

It is notable that the student took the video while leaning over a railing, with only his hands for support. The experiment grew organically after a group of students finished a game of eight ball at our department's winter holiday party at a colleague's home.  Only a few videos where taken, and the only careful measurement was that of the dimensions of the pool table by a tape measure provided by the host.  It is astonishing that such a quick and simple experiment can yield such a trove of information, as we show below.  This illustrates how access to a smart phone and Video Extractor provides many opportunities for a student to do experiments when an interesting opportunity arises during their daily routine.

Figure \ref{ref:Frame} shows one frame in the video, just before the cue stick strikes the cue ball.  The distance between cushions was measured to be 44.0" in width and 88.0" in length, which we converted to 1.12 m by 2.24 m.  In video extractor, the axes were centered on the left near pocket relative to the player and drawn to run along the two adjacent cushions along the full length and width of the table.  The scale was then set to 1.12 m by 2.24 m.

The cursor was placed on the middle of the ball, single clicked, then repositioned on the middle of the ball repeatedly for at least 5 repetitions of a single click before double clicking to advance the frame.  This allowed a determination of the uncertainty from the standard deviation of the multiple measurements.

\section{Results and discussions}

Origin's video extractor is used to determine the (x,y) coordinates of the cue ball and the eight ball as a function of time.  From this, the distance travelled
\begin{align} \label{eq:distance}
d = \sqrt{x^2 + y^2}
\end{align}
as a function of time  is determined.  The uncertainty is propagated using the quadrature
\begin{align} \label{eq:uncertainty}
\Delta d = \sqrt{(x \Delta x)^2 + (y \Delta y)^2} / d .
\end{align}

\begin{figure}[h]
    \centering
       \includegraphics{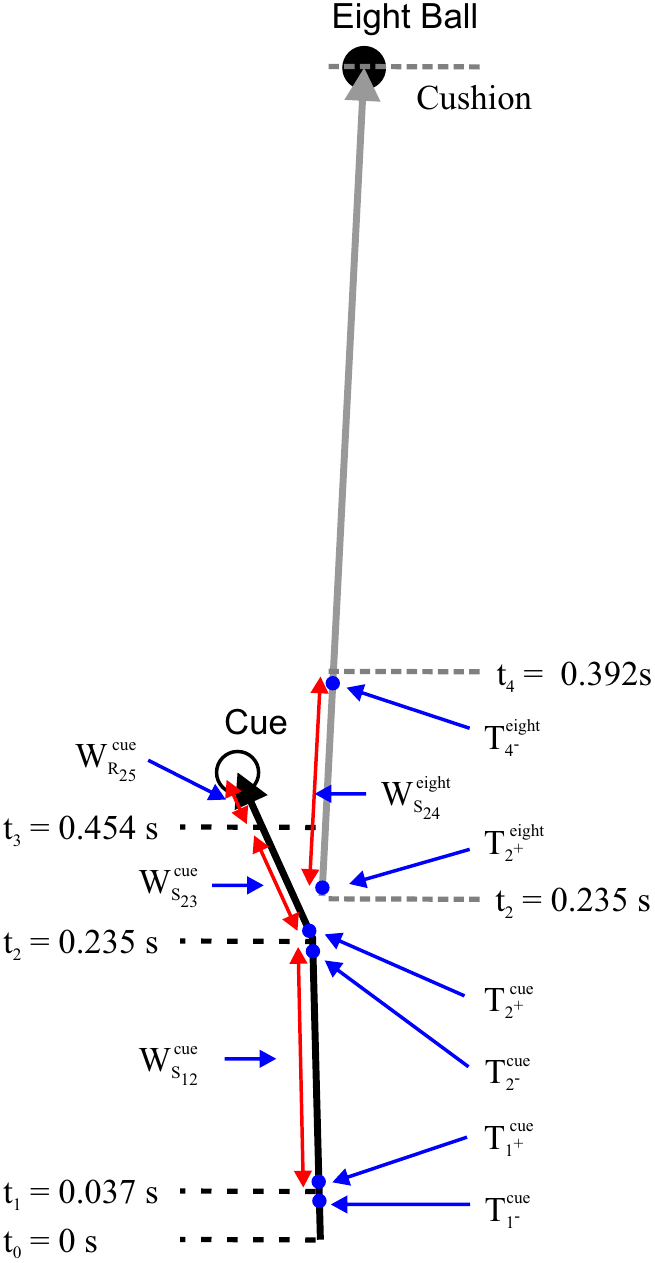}
    \caption{The trajectories of the cue ball and the eight ball at key times of various events.  At $t_0 =0$ the video starts; at $t_1=0.037 \, \text{s}$ the cue hits the cue ball; at $t_2 = 0.235 \, \text{s}$ the cue ball collides with the eight ball; at $t_3=0.454 \, \text{s}$ the cue ball stops sliding and starts to roll; and at $t_4=0.392 \, \text{s}$ the eight ball stops sliding and starts to roll.  The various labels $T$ are the translational kinetic energies at the various events shown for each ball.  The various labels $W$ represent the work done by friction between the various events shown. }
    \label{fig:Trajectories}
\end{figure}

Figure \ref{fig:Trajectories} summarizes the trajectories of the cue ball and the eight ball, and labels the kinetic energies before and after various events (such as collisions).  Also labeled is the work done by different sources of friction in various regions.

\subsection{Equations of Motion}

This section determines the equations of motion of the cue ball and the eight ball between events, as labelled by the blue points in Figure \ref{fig:Trajectories}.  We will show that a fit of the position as a function of time fits a quadratic function for each ball and in each region.  The quadratic coefficient gives the acceleration.  Since the table is level, the only force acting between events (collisions) is friction.  We find two distinct accelerations, which we associate with rolling friction (the smaller force) and sliding friction (the larger force).  For each ball, we calculate the coefficient of friction and compare it with typically-measured values.

\subsubsection{The Cue Ball's Motion}

\begin{figure}[h!]
    \centering
       \includegraphics{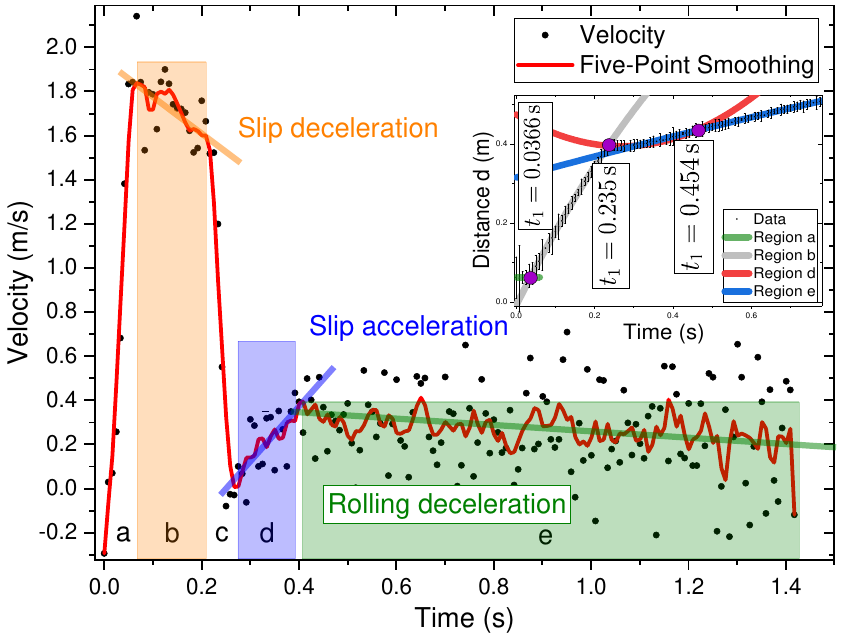}
    \caption{The velocity of a cue ball as a function of time (points) and five-point smoothing (red curve).  The cue ball is struck in the region labelled $a$ and hits the eight ball in Region $c$.  The other regions correspond to the cue ball sliding ($b$ and $d$) and rolling ($e$).  The inset shows a plot of the distance as a function of time and piecewise quadratic fits, with the intersections shown as large purple points.}
    \label{fig:CueVel}
\end{figure}
Figure \ref{fig:CueVel} shows the velocity of the cue ball as a function of time, which is determined from Origin's derivative function of the distance travelled.  Because the derivative is calculated from differences between adjacent positions, the noise is amplified.  In our experiment, when the cue ball is traveling at its highest speed, it traverses about 5 pixels per frame.  Because digitization gives the coordinate of the nearest pixel, the uncertainty in the position is about half a pixel.  The difference between two positions will thus have an uncertainty spread of about 1.4 pixels, or almost 30\% for a 5 pixel separation.  That is the approximate scatter of the velocity data in Region $b$, where the cue ball's speed is the highest.  At later times, when the speed is lower, the noise is even more amplified.   Five-point smoothing is applied to reduce scatter to help visualize the trends.  Also shown are straight line segments that follow the data.  {\em Such velocity plots are used solely as a visual aid to interpreting the processes taking place, and not for determining fit parameters.}  Only the more precise position data as plotted in Figure \ref{ref:CueBall} are used to get the parameters from quadratic fits.

The letters at the bottom of the plot label five distinct regions.  Region $a$ corresponds to the time of the cue ball's acceleration when it is in contact with the cue.

The cue ball travels freely in Region $b$ but is decelerating due to slipping friction.  We base this assumption on the fact that the observed deceleration is too high for rolling friction alone to be the cause (we test this hypothesis later).

Region $c$ is the time when the cue ball comes in contact with the eight ball, causing the cue ball to decelerate as it transfers energy and momentum to the eight ball.

In Region $d$ the cue ball accelerates.  This indicates that the ball is rotating at a higher angular velocity than the rolling angular velocity.  This phenomena can be simply explained by considering a rolling ball after a collision.  The speed decreases abruptly due to the impulse but the ball continues to spin at the same rate because the torque acts for a short time (described later).  The ball then accelerates as angular energy is transferred to translational energy.

In Region $e$ the ball is no longer slipping, and it is decelerating due to rolling friction.

Figure \ref{ref:CueBall} shows the distance travelled as a function of time.  We use Figure \ref{fig:CueVel} to approximately determine the three regions to fit, corresponding first to slipping deceleration after the cue strikes the cue ball, slipping acceleration after the cue strikes the eight ball followed by rolling deceleration.

A more objective method for determining the times of the events is to fit each region separately to the quadratic function
\begin{align}\label{eq:dForAccelration}
d = d_0 + v_0 t + \frac {1} {2} a t^2 ,
\end{align}
where to first approximation the regions are determined ``by eye."  Next, using the fits parameters, the intersections between the fit curves are determined analytically.  The curves in Figure \ref{ref:CueBall} show these fits.  The inset in Figure \ref{fig:CueVel} shows the data and the fit functions, which are plotted beyond the domain of the data to make the intersection points more obvious.  The analytically-determined intersection times are shown as purple dots.

The flat light green region in the inset of Figure \ref{fig:CueVel} represents the stationary ball before it is struck by the cue stick, depicted by the purple dot at the intersection between the light green and gray curves.  The ball follows the gray trajectory until it hits the eight ball, depicted by the purple dot at the intersection of the gray and red curves.  The cue ball then follows the red curve until it stops slipping and rolls, which corresponds to the blue curve.  The boxes in Figure \ref{ref:CueBall} highlight the fit parameters used to generate the curves and Table \ref{tab:QuadParamCue} summarizes the values of $d$, $v$ and $a$ at the start and end times of each region as well as the analytically-determined intersection times.

\begin{table}\footnotesize
        \centering
        \caption{Distance, velocity and acceleration of the cue ball determined from quadratic fits at the specified time intervals.\label{tab:QuadParamCue}}
        \label{multiprogram}
        \begin{tabular}{c c c c}
            & & & \\
              \multicolumn{4}{l}{{\bf Region $a$}: $0\, \mbox{s} < t < 0.0366 \, \mbox{s} $ -- Cue Stationary} \\
            \hline
            \hline
            $t \, (\mbox{s}) $ & $d \, (\mbox{m}) $ & $v \, (\mbox{m}/\mbox{s})$ & $a \, (\mbox{m} / \mbox{s}^2)$ \\
            \hline
            0 & 0.06314 $\pm$ 0.00020 & 0 & 0 \\
            0.0366 & 0.06314 $\pm$ 0.00020 & 0 & 0 \\
            \hline
            \hline
             & & & \\[+1em]
             \multicolumn{4}{l}{{\bf Region $b$}: $0.0366 \, \mbox{s} < t < 0.235 \, \mbox{s} $ -- Cue Slip Deceleration}\\
            \hline
            \hline
            $t \, (\mbox{s}) $ & $d \, (\mbox{m}) $ & $v \, (\mbox{m}/\mbox{s})$ & $a \, (\mbox{m} / \mbox{s}^2)$ \\
            \hline
            0.0366 & 0.0631 $\pm$ 0.0031 & 1.894 $\pm$ 0.040 & -2.21 $\pm$ 0.26 \\
            0.235 & 0.396 $\pm$ 0.012 & 1.455 $\pm$ 0.073 & -2.21 $\pm$ 0.26 \\
            \hline
            \hline
             & & & \\
            \multicolumn{4}{c}{ --- Cue Collides with Eight Ball ---} \\
             & & & \\
            \multicolumn{4}{l}{{\bf Region $d$}: $0.235 \, \mbox{s} < t < 0.454 \, \mbox{s} $  -- Cue slip Acceleration}\\
            \hline
            \hline
            $t \, (\mbox{s}) $ & $d \, (\mbox{m}) $ & $v \, (\mbox{m}/\mbox{s})$ & $a \, (\mbox{m} / \mbox{s}^2)$ \\
            \hline
            0.235 & 0.396 $\pm$ 0.036 & -0.080 $\pm$ 0.015 & +2.23 $\pm$ 0.38 \\
            0.454 & 0.432 $\pm$ 0.070 & 0.41 $\pm$ 0.21 & +2.23 $\pm$ 0.38 \\
            \hline
            \hline
            & & & \\[+1em]
            \multicolumn{4}{l}{{\bf Region $e$}: $0.454 \, \mbox{s} < t < \infty $  -- Cue Rolling Deceleration}\\
            \hline
            \hline
            $t \, (\mbox{s}) $ & $d \, (\mbox{m}) $ & $v \, (\mbox{m}/\mbox{s})$ & $a \, (\mbox{m} / \mbox{s}^2)$ \\
            \hline
            0.454 & 0.4325 $\pm$ 0.0027 & 0.2445 $\pm$ 0.0065 & -0.0594 $\pm$ 0.0064 \\
            4.57 & 0.9358 $\pm$ 0.072 & 0 & -0.0594 $\pm$ 0.0064 \\
            \hline
            \hline
        \end{tabular}
    \end{table}

\subsubsection{Coefficient of Sliding/Rolling Friction of the Cue Ball}

\begin{figure}[h!]
    \centering
       \includegraphics{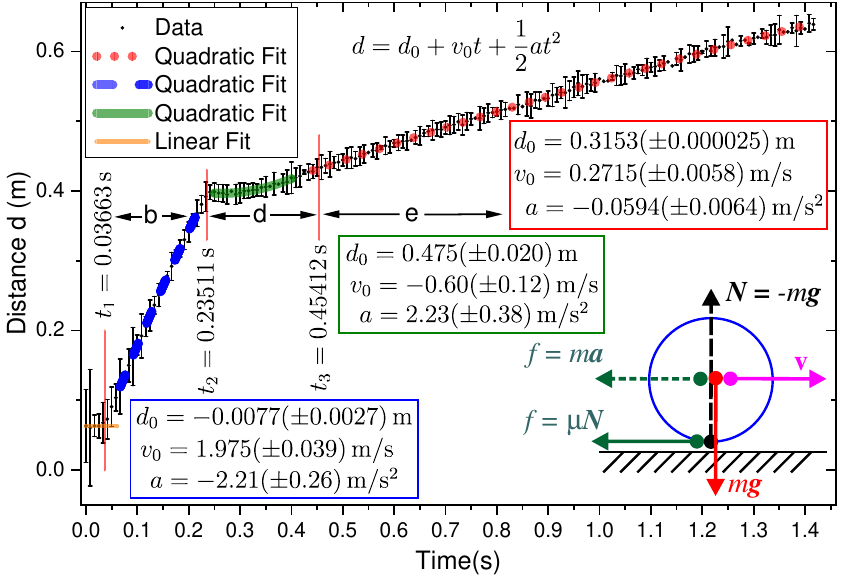}
    \caption{The distance travelled by the cue ball as a function of time after impact with the cue and continuing after colliding with the eight ball.  The points show the data and the curves are piecewise quadratic fits to the various regions.  The fit parameters are shown inside boxes near each region and of the same color as the corresponding plot.  The bottom right inset shows the free body diagram of the cue ball while slipping.}
    \label{ref:CueBall}
\end{figure}
We start by calculating the coefficient of sliding friction and rolling friction\cite{sooda96.01,knopp19.01}  The coefficient of sliding friction is given by the ratio of the frictional force $f$ to the normal force $N$, or
\begin{align} \label{eq:SlidingFric}
\mu_s = \frac {f} {N} = \frac {m a} {m g} = \frac {a} {g} ,
\end{align}
where $m$ is the mass of the cue ball and the frictional force is determined from the ball's acceleration as shown in the free body diagram in Figure \ref{ref:CueBall}.  Rolling friction is calculated in the same way.  During slip deceleration, this yields a coefficient of friction  $\mu_s^\text{cue} = 0.226 (\pm 0.027)$.  The slip acceleration region gives $\mu_s^\text{cue} = 0.228 (\pm 0.039)$.  Note that the coefficient of friction is independent of speed, as we would expect for low speeds,\cite{Xu07.01} and the same value is obtained both during slip deceleration and acceleration.  The values compare favorably with the range observed of 0.15 to 0.40, with 0.20 being the most common value.\cite{alcia20.01} The coefficient of rolling friction is found to be given by $\mu_r^\text{cue} = 0.00606 (\pm 0.00065)$, which falls in the most-often observed range of~0.005 to~0.015.

\begin{figure}[h!]
    \centering
       \includegraphics[width= 3.4 in, height = 2.5 in]{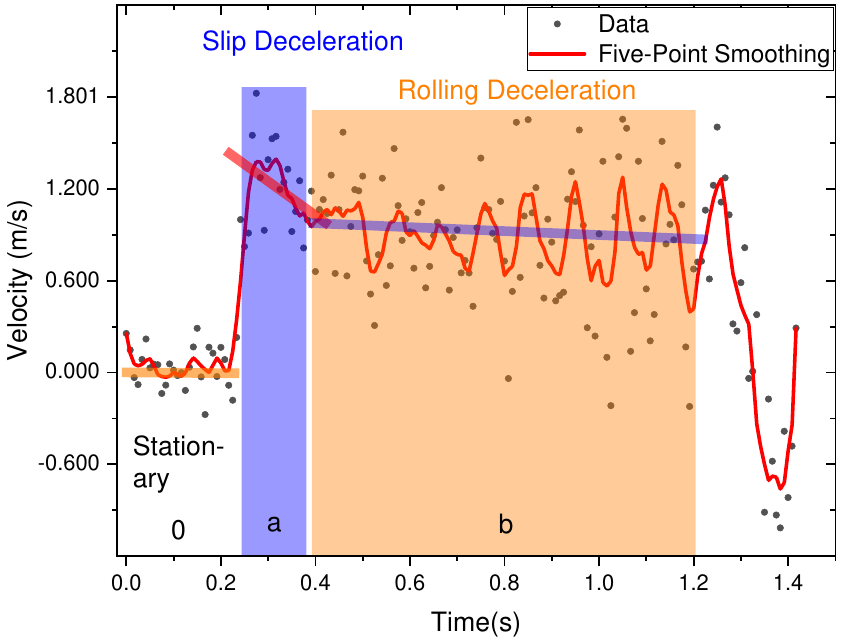}
    \caption{The velocity of the eight ball as a function of time (points) and five-point smoothing (red curve).  The eight ball is struck by the cue ball in the region labelled ``0.''  The other regions correspond to the eight ball slipping (Region $a$) and rolling (Region $b$).  The oscillations are an artifact of difficulties associated with visually determining the center of a dark object on a dark background, as discussed in the text.}
    \label{fig:EightVel}
\end{figure}

\subsubsection{The Eight Ball's Motion}

Figure \ref{fig:EightVel} shows the velocity of the eight ball as a function of time, which is determined from Origin's derivative function of the distance traveled, as we did for the cue ball in Figure \ref{fig:CueVel}. There are three distinct regions in Figure \ref{fig:EightVel}: (1) Region $0$ where the eight ball is stationary, (2) Region $a$ where the eight ball is decelerating due to sliding friction, and (3) Region $b$ where the eight ball is no longer slipping and is decelerating due to rolling friction.

\begin{figure}[h!]
    \centering
    \includegraphics[width = 3.4 in, height = 2.5 in]{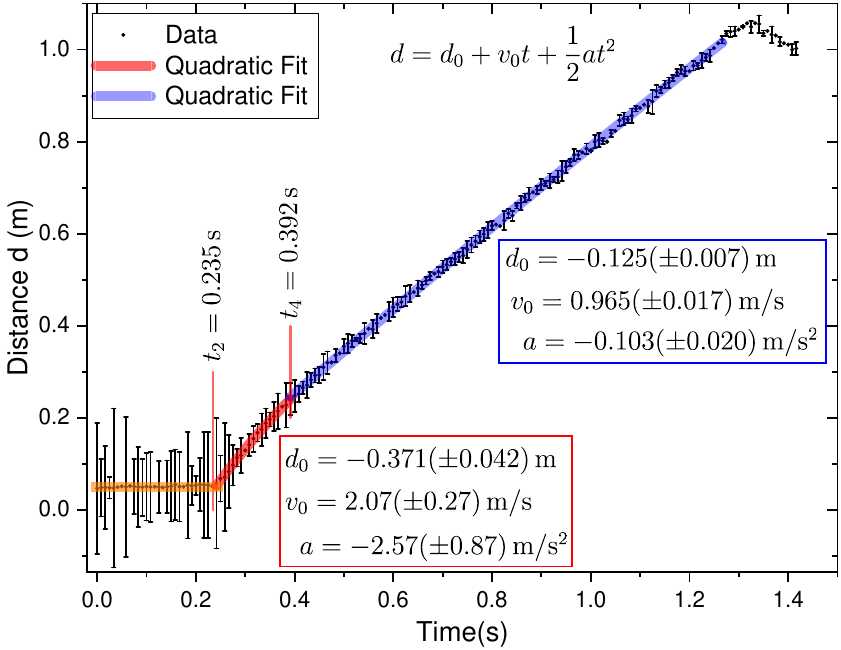}
    \caption{The distance travelled by the eight ball as a function of time after impact with the cue ball and continuing after the collision.  The points show the data and the curves are quadratic fits to the various regions.  The fit parameters are shown near each region.}
    \label{ref:EightBall}
\end{figure}

Figure \ref{ref:EightBall} shows the distance travelled by the eight ball as a function of time. The quadratic function given by Equation \ref{eq:dForAccelration} is used to fit each region in Figure \ref{ref:EightBall} for the eight ball as we did for the cue ball in Figure \ref{ref:CueBall}.  The orange fit line in Figure \ref{ref:EightBall} corresponds to the eight ball being stationary until struck by the cue ball at the intersection between the orange and red curves. The eight ball starts slipping and follows the red trajectory until it stops slipping and starts rolling at the intersection of the red and blue curves. The eight ball then takes the blue trajectory until it hits the cushion. The boxes in Figure \ref{ref:EightBall} show the fit parameters used to generate the curves and Table \ref{tab:QuadParamEight} summarizes the values of $d$, $v$ and $a$ at the start and end times of each region.

\subsubsection{Coefficient of Sliding/Rolling Friction of the Eight Ball}

The coefficient of sliding and rolling friction are calculated using the eight ball's acceleration data and Equation \ref{eq:SlidingFric}.  During slip deceleration, the coefficient of sliding friction in Region $a$ is $\mu_s = 0.262(\pm 0.088)$, which falls within the typical range of $0.15$ to $0.40$ for sliding friction on a pool table\cite{alcia20.01} and agrees with the value obtained for the cue ball within experimental uncertainty.  The coefficient of rolling friction is given by $\mu_r^\text{eight} = 0.011(\pm 0.002)$ and falls into the typically-observed range of $0.005$ and $0.015$, but does not agree with  $\mu_r^\text{eight} = 0.00606 (\pm 0.00065)$, the value for the cue ball.  The difference might be due to the cue ball and eight ball rolling on different parts of the table where the felt's properties are different or in the properties of the balls.

\begin{table}\footnotesize
        \centering
       \caption{Distance, velocity and acceleration of the eight ball determined from quadratic fits at the times specified in each region.\label{tab:QuadParamEight}}
        \begin{tabular}{c c c c}
            \\
             \multicolumn{4}{l}{{\bf Region 0}: $0\, \mbox{s} < t < 0.235 \, \mbox{s} $ -- Eight Ball Stationary}\\
            \hline
            \hline
            $t \, (\mbox{s}) $ & $d \, (\mbox{m}) $ & $v \, (\mbox{m}/\mbox{s})$ & $a \, (\mbox{m} / \mbox{s}^2)$ \\
            \hline
            0 & 0.044 $\pm$ 0.081 & 0 & 0 \\
            0.235 & 0.044 $\pm$ 0.081 & 0 & 0 \\
            \hline
            \hline
             & & & \\
            \multicolumn{4}{c}{ --- Cue Collides with Eight Ball ---} \\
             & & & \\
             \multicolumn{4}{l}{{\bf Region $a$}: $0.235 \, \mbox{s} < t < 0.392 \, \mbox{s} $ -- Slip Deceleration}\\
            \hline
            \hline
            $t \, (\mbox{s}) $ & $d \, (\mbox{m}) $ & $v \, (\mbox{m}/\mbox{s})$ & $a \, (\mbox{m} / \mbox{s}^2)$ \\
            \hline
            0.235 & 0.044 $\pm$ 0.081 & 1.46 $\pm$ 0.34 & -2.57 $\pm$ 0.87 \\
            0.392 & 0.24 $\pm$ 0.13 & 1.06 $\pm$ 0.44 & -2.57 $\pm$ 0.87 \\
            \hline
            \hline
             & & & \\[+1em]
             \multicolumn{4}{l}{{\bf Region $b$}: $0.392 \, \mbox{s} < t < 1.2  \, \mbox{s} $ -- Rolling Deceleration }\\
            \hline
            \hline
            $t \, (\mbox{s}) $ & $d \, (\mbox{m}) $ & $v \, (\mbox{m}/\mbox{s})$ & $a \, (\mbox{m} / \mbox{s}^2)$ \\
            \hline
            0.392 & 0.245 $\pm$ 0.009 & 0.924 $\pm$ 0.019 & -0.103 $\pm$ 0.020 \\
            1.2 & 0.951 $\pm$ 0.026 & 0.841 $\pm$ 0.030 & -0.103 $\pm$ 0.020 \\
            \hline
            \hline
        \end{tabular}
    	\label{table: eightFitValues}
    \end{table}

\subsection{Angular Momentum and Frictional Work/Torque}

This section uses the parameters found from the quadratic fits to the position data to calculate the  frictional torque and frictional work done on each ball and the resulting change in angular momentum.  Since it is difficult to observe rotation of the balls in the video, their angular velocity cannot be easily determined, so we will calculate it using Newton's second law, which relates the torque to the change in angular momentum.

\subsubsection{Cue Ball}\label{sec:CueAndMom}

\begin{figure}[h!]
    \centering
    \includegraphics{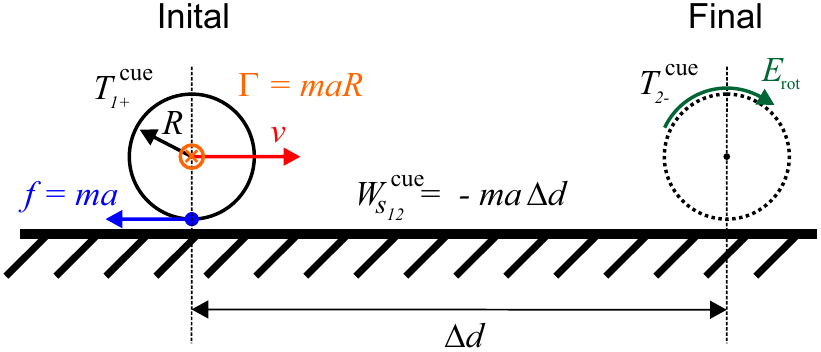}
    \caption{A free-body diagram of the cue ball just after being struck by the cue stick (solid circle) and just before colliding with the eight ball (dashed circle).}
    \label{fig:FBD-Slipping}
\end{figure}

Figure \ref{fig:FBD-Slipping} shows a free-body diagram of the cue ball just after being struck by the cue stick (initial) and just before colliding with the eight ball (final).  It is assumed to have no angular velocity just after being struck by the eight ball.  We argue this to be so from Newton's second law, which gives the change of the
angular momentum $\Delta L$ in time interval $\Delta t$ to be given by
\begin{align}\label{eq:AngMom}
\Delta L = \Gamma \Delta t = R f \Delta t,
\end{align}
where $\Gamma$ is the frictional torque and $R= 2.86 \times 10^{-2} \, \mbox{m}$, the radius of the cue ball.  $\Delta L$ will be much smaller than the angular momentum gained as the ball is slipping since $\Delta t$ during the time the cue ball is in contact with the cue stick is short compared with the time the cue ball is sliding.  The cue ball slides for about 0.2$\,$s before it hits the eight ball, which is certainly much longer than the time the cue stick is in contact with the cue ball.

Assuming that the initial angular momentum is negligible, the change in the angular momentum is the total angular moment at the time the cue ball strikes the eight ball.  At this time, with $\Delta t = 0.198 \, \mbox{s}$, Equation \ref{eq:AngMom} yields
\begin{align} \label{eq:AngMomValue}
L^\text{cue} = 2.13 (\pm 0.25) \times 10^{-3} \, \mbox{J} \cdot \mbox{s}
\end{align}
in the horizontal direction.

The deceleration of the cue ball in Region $b$ of Figure \ref{fig:CueVel} is uniform and large so it appears that it is undergoing slip deceleration the whole time between the cue ball being struck by the cue stick and colliding with the eight ball.  To check if the cue is slipping and not rolling, we use the angular momentum from Equation \ref{eq:AngMomValue} to determine the speed $v_R$ if it were rolling at the time the cue ball collides with the eight ball, which yields
\begin{align} \label{eq:vRoll}
v_R^\text{cue} = \frac {R^\text{cue} L^\text{cue}} {I^\text{cue}} = 1.10 (\pm 0.13) \, \mbox{m} / \mbox{s} ,
\end{align}
with the moment of inertia given by
\begin{align}\label{eq:MomIner}
I^\text{cue} = \frac {2} {5} m^\text{cue} (R^\text{cue})^2 = 5.56 \times 10^{-5} \, \mbox{kg} \cdot \mbox{m}^2.
\end{align}
The physical parameters are summarized in Table \ref{tab:BallProps}.

Table \ref{tab:QuadParamCue} gives a measured speed of $v^\text{cue} = 1.455 (\pm 0.073 ) \, \mbox{m} / \mbox{s}$ at the time the cue strikes the eight ball, which is larger than the rolling speed given by Equation \ref{eq:vRoll}.  Thus, the cue ball is still slipping and far from rolling when colliding with the eight ball, as expected from the insights offered by Figure \ref{fig:CueVel}.

\begin{table}\footnotesize
        \centering
        \caption{The cue and eight ball's properties and sliding friction.\label{tab:BallProps}}
        \begin{tabular}{r c c}
            \hline
            Property & Cue & Eight\\
            \hline
            \hline
            Mass & $ 0.17 \, kg$ & $ 0.16 \, kg$ \\
            Radius & $0.0286 \, m$ & $0.0286 \, m$ \\
            Moment of Inertia & $5.56 \times 10^{-5}\,kg \cdot m^2$ & $5.23 \times 10^{-5}\,kg \cdot m^2$ \\
            Coefficient of Friction & $0.226 (\pm 0.027)$ & $0.262(\pm 0.088)$ \\
            Sliding Friction Force & $0.376 (\pm 0.044) \, N$ & $0.41 (\pm 0.14) \, N $\\
            \hline
        \end{tabular}
    \end{table}

\subsubsection{Eight Ball}

The calculations for the eight ball's motion mirror that of the cue ball, except that the eight ball is struck by the cue ball rather than the cue stick.  Applying Equation \ref{eq:AngMom} under the assumption that the eight ball's initial angular momentum after the collision with the cue ball vanishes, and using the values in Table \ref{tab:BallProps} and the slip deceleration time from Table \ref{tab:QuadParamEight} of $\Delta t = 0.157 \, s$ yields
\begin{align} \label{eq:AngMomEight}
L^\text{eight} = 1.84 (\pm 0.63) \times 10^{-3} \, \mbox{J} \cdot \mbox{s} .
\end{align}

As an exercise, the student should apply the arguments from Section \ref{sec:CueAndMom} to determine if the eight ball is indeed rolling at $t = 0.392 \, \text{s}$.

\subsection{Work-Energy Theorem}\label{sec:CueEnergyCons}

Here we apply the work-energy theorem to various parts of each ball's motion. First we start with a general introduction.

Consider the free-body diagram in Figure \ref{fig:FBD-Slipping}.  Note that the diagram is drawn for the cue ball but the results are general.  The only force acting on the ball is that of slipping friction, which acts on the contact point of the ball with the table, as shown.  This force decelerates the center of mass' linear motion and the torque it applies increases the angular momentum.  The work-energy theorem applied to the translational motion is
\begin{align}\label{eq:Work-Energy-Theorem-Trans}
E_f^\text{trans} - E_i^\text{trans} = \int_i^f {\bf F} ({\bf r}) \cdot \,  d {\bf r}
\end{align}
and applied to the rotational motion is
\begin{align}\label{eq:Work-Energy-Theorem-Rotate}
E_f^\text{rot} - E_i^\text{rot} = \int_i^f {\bf \Gamma} ({\bf r}) \cdot \,  d {\boldsymbol \phi}
\end{align}

Since the frictional force is constant during slipping, Equation \ref{eq:Work-Energy-Theorem-Trans} becomes
\begin{align}\label{eq:Work-Energy-Theorem-Trans-Fric}
E_f^\text{trans} - E_i^\text{trans} = {\bf F}  \cdot \int_i^f  \,  d {\bf r} \rightarrow \Delta T = F  \cdot \Delta d ,
\end{align}
where we define $\Delta T = E_f^\text{trans} - E_i^\text{trans}$, and when applied to the rotational motion for constant torque Equation \ref{eq:Work-Energy-Theorem-Rotate} becomes
\begin{align}\label{eq:Work-Energy-Theorem-Rotate-Fric}
E_f^\text{rot} - E_i^\text{rot}  =  {\bf \Gamma}  \cdot \int_i^f  \,  d {\boldsymbol \phi} \rightarrow \Delta E_{rot} = \Gamma   \Delta \phi ,
\end{align}
where $\Delta d$ is the distance the ball has traveled and $\Delta \phi$ its angle of rotation.

\subsubsection{Cue Ball}

The initial energy of the cue ball just after it is struck by the cue stick is all carried as translational kinetic energy.  As the cue slides, it decelerates and the translational kinetic energy is lost to rotational energy and dissipated into heat by friction.  This section checks the work energy theorem for both the rotational and translational parts.

First we evaluate Equation \ref{eq:Work-Energy-Theorem-Trans-Fric} to verify the translational work-energy theorem.  The velocity of the cue ball just after it is struck by the cue stick is given in Table \ref{tab:QuadParamCue} at the start of Region $b$, which gives the translational kinetic energy
\begin{align}\label{eq:InitialE-Cue}
E_i^\text{cue} = T_{1^+}^\text{cue} = \frac {1} {2} m v_1^2 = 0.305 (\pm 0.013) \, \mbox{J},
\end{align}
which is also the initial energy.  The mass uncertainties are ignored because they are negligible compared with the uncertainties of all other quantities.  The slipping force is $f^\text{cue} = m a = 0.376 (\pm 0.044) \, \mbox{N}$ leading to the frictional work
\begin{align}\label{eq:friction}
W_{s_{12}}^\text{cue} = - f^\text{cue} \Delta d = - m a \Delta d = - 0.125 (\pm 0.015) \, \mbox{J},
\end{align}
where $\Delta d = 0.333 (\pm 0.012) \, \mbox{m}$ and the sign is negative because the direction of the frictional force is opposite to the displacement.

The speed of the cue ball just before it hits the eight ball is $v_2 = 1.455 (\pm 0.073) \, \mbox{m/s}$, giving a translation kinetic energy
\begin{align}\label{eq:CueKEBeforeEightHit}
T_{2^-}^\text{cue} \equiv T_2 (t = \lim_{\epsilon \rightarrow 0} (t_2 - \epsilon) ) = 0.1799 (\pm 0.0090) \, \mbox{J} .
\end{align}

Now let's check if the Work-Energy Theorem given by Equation \ref{eq:Work-Energy-Theorem-Trans-Fric} holds as the cue ball is slipping.  Using Equations \ref{eq:InitialE-Cue} and \ref{eq:CueKEBeforeEightHit}, we get
\begin{align}\label{eq:Econserve}
E_f^\text{cue} - E_i^\text{cue} =  T_{2^-}^\text{cue} - T_{1^+}^\text{cue} = -0.125(\pm 0.016) \, \text{J}.
\end{align}
The values in Equations \ref{eq:friction} and \ref{eq:Econserve} agree within experimental uncertainties.

We cannot evaluate Equation \ref{eq:Work-Energy-Theorem-Rotate-Fric} to verify the rotational part of the Work-Energy Theorem because we cannot observe the rotation of the cue ball.  However, we can calculate the rotational energy of the cue ball just when it strikes the eight ball, yielding
\begin{align}\label{eq:RotEnergy}
E_\text{rot}^\text{cue} = \frac {L^2} {2I} = 0.0407 (\pm 0.0048) \, \mbox{J},
\end{align}
where we have used the angular momentum calculated from Equations \ref{eq:AngMom} and \ref{eq:AngMomValue}.

Since the cue ball initially has negligible angular momentum, the rotational energy yields $E_f^\text{rot} - E_i^\text{rot} = E_\text{rot}^\text{cue}$.  We can then use Equation \ref{eq:Work-Energy-Theorem-Rotate-Fric} in reverse to get $\Delta \phi$  with $\Gamma = f R$, which yields
\begin{align}\label{eq:RotAngle}
\Delta \phi = \frac {E_\text{rot}^\text{cue}} {f^\text{cue}R}.
\end{align}
Substituting the numbers into Equation \ref{eq:RotAngle} does not give us any interesting information unless we had a measurement of the angular motion for comparison.

Finally, we can combine both the angular and translational motion into a more general Work-Energy Theorem by adding Equations \ref{eq:Work-Energy-Theorem-Trans-Fric} and \ref{eq:Work-Energy-Theorem-Rotate-Fric}, yielding
\begin{align}\label{eq:EconserveTotal}
E_f^\text{tot} - E_i^\text{tot}  = T_{2^-}^\text{cue} - T_{1^+}^\text{cue} + E_\text{rot}^\text{cue} = F \cdot \Delta d + \Gamma \Delta \phi .
\end{align}
Since Equations \ref{eq:Work-Energy-Theorem-Trans-Fric} and \ref{eq:Work-Energy-Theorem-Rotate-Fric} each hold, so will the sum.

As an exercise, the student should apply the Work-Energy Theorem to the cue ball while it is rolling after the collision.

\subsubsection{Eight Ball}

Next we consider the motion of the eight ball after it is struck by the cue ball.  As for the cue ball, we assume that the eight ball's angular velocity is small just after it is struck.

The eight ball slides a distance of $\Delta d = 0.188(\pm 0.157) \, \mbox{m}$ after being struck by the cue ball.  With slipping force $f^\text{eight} =  0.41(\pm 0.14) \, \mbox{N}$, the frictional work is
\begin{align}\label{eq:EightBallSlippingWork}
W_{s_{24}}^\text{eight} = 0.081(\pm 0.070) \, \mbox{J} .
\end{align}
The change in kinetic energy is
\begin{align}\label{eq:EconserveEight}
E_f^\text{eight} - E_i^\text{eight} =  T_{4^-}^\text{eight} - T_{2^+}^\text{eight} = -0.08(\pm 0.11) \, \text{J},
\end{align}
where we have used $T_{2^+}^\text{eight} = \frac{1}{2}mv^2 (t = 0.236^+) = 0.171(\pm 0.057) \mbox{J}$ and $T_{4^-}^\text{eight} = \frac{1}{2}mv^2 (t = 0.392^-) = 0.090(\pm 0.075) \mbox{J}$.  Equations \ref{eq:EightBallSlippingWork} and \ref{eq:EconserveEight} agree, so the work-energy theorem is obeyed.

It may seem somewhat suspicious that the agreement is so good even when the uncertainties are so large.  This arises from the fact that the energies and work are determined from the equations of motion that were applied in getting the velocities, distances and frictional forces.  Since the Work-Energy Theorem is identical to Newton's laws, using the parameters that we got from Newton's laws should yield the same result as does the work-energy theorem.  This illustrates numerically how the two approaches are indeed identical.

\subsection{Energy Conservation}

The physics of a sliding and rolling ball demonstrates many physical principles but is complex enough for a typical student to have difficulties in sorting these out.  Then, it is best to consider limiting cases, as follows.

First, when an object slides on a frictionless surface and collides elastically with other objects, the total energy of the system is constant as implied by the work-energy theorem because there are no external forces.  Next consider a block sliding on a surface with coefficient of friction $\mu$, which according to the Work-Energy Theorem given by Equation \ref{eq:Work-Energy-Theorem-Trans-Fric} gives
\begin{align}\label{eq:W-E-Block}
E_f = E_i - \mu N \Delta d ,
\end{align}
so the kinetic energy decreases in proportion to the distance travelled.  This lost energy is dissipated as heat, etc.~as usually explained with friction.

The case of the sliding ball falls in the intermediate regime where some of the translational energy is converted to rotational energy while some of the energy is lost to dissipative process associated with friction.  If two balls collide, linear and rotational kinetic energy will be conserved as long as the balls are in contact over a short enough time that the righthand sides of Equations \ref{eq:Work-Energy-Theorem-Trans-Fric} and \ref{eq:Work-Energy-Theorem-Rotate-Fric} vanish, which we have argued above is the case in our experiments.

Note that students often believe that frictional forces lead only to energy dissipation, which is true for a sliding block.  However, this paper illustrates that frictional forces can also be responsible for converting one type of kinetic energy into another.

Below we start by calculating the energy lost due to dissipative forces while slipping then consider energy conservation in the collision process.

\subsubsection{Energy Dissipated in a Slipping Ball}

When the ball slides on a surface with friction, some translational kinetic energy is converted to rotational kinetic energy and some energy is dissipated.  The former has already been calculated above.  Here we calculate the energy lost to dissipative processes.

For the Cue Ball, the energy dissipated is simply given by the total change in kinetic energy, or
\begin{align}\label{eq:Dissipated}
E_D^\text{cue} = E_f^\text{cue} - E_i^\text{cue}  = T_{2^-}^\text{cue} - T_{1^+}^\text{cue} + E_\text{rot}^\text{cue}  .
\end{align}
Substituting the values from Equations \ref{eq:RotEnergy} and \ref{eq:Econserve} into Equation \ref{eq:Dissipated} yields
\begin{align}\label{eq:DissipatedCue}
E_D^\text{cue} =  0.084(\pm 0.017) \, J  .
\end{align}
Thus, the energy lost to dissipative processes is a large fraction of the total energy of the system and larger than the rotational energy gained.

The student should do the following exercises:
\begin{enumerate}

\item Follow the line of reasoning above to determine the dissipative energy loss of the eight ball while slipping.

\item Calculate the energy dissipated as the cue ball accelerates after colliding with the eight ball.

\end{enumerate}

\subsubsection{Energy Conservation in the Collision}

This section determines how much of the energy lost by the cue ball is transferred to the eight ball.  If they balance, the collision is elastic.

We start by determining the energy lost by the cue ball upon colliding with the eight ball.  Since no torque is applied in a collision where the impulse is directed at the center of mass and the impulse is short enough in duration for frictional torque to not have acted, the angular velocity of the cue ball does not change appreciably while it is in contact with the eight ball, and the change in the total energy of the cue ball is the change in translational kinetic energy given by
\begin{align}\label{eq:ChangeKinCueEight}
\Delta T^\text{cue} &= T_{2^+}^\text{cue} - T_{2^-}^\text{cue} \nonumber \\
&= \frac {1} {2} (0.17 \, \mbox{kg}) (-0.080(\pm 0.015) \, \mbox{m/s} )^2 \nonumber \\
&- \frac {1} {2} (0.17 \, \mbox{kg})  (1.455(\pm 0.073) \, \mbox{m/s})^2 \nonumber \\
&= -0.180(\pm 0.033) \, \mbox{J} ,
\end{align}
where we have used the velocities in Table \ref{tab:QuadParamCue}.  Note that the fact that the cue ball's velocity is small right after the collision is consistent with the video, which shows the cue ball to appear to stop immediately after the collision, then accelerates as rotational energy is converted to translation kinetic energy.

The kinetic energy of the eight ball right after the collision is given by
\begin{align}\label{eq:EightBallInitialKE}
\Delta T^\text{eight} = T_{2^+}^\text{eight} &= \frac {1} {2} m [v(0.235^+ \, \mbox{s})]^2 \nonumber \\
&= 0.170 (\pm 0.040 ) \, \mbox{m/s},
\end{align}
where we have used the velocity from Table \ref{table: eightFitValues}.  Note that this is the change in kinetic energy since the eight ball starts from rest.  Assuming that rotational energy has not changed in the collision, if the collision were elastic then the energy lost by the cue ball given by Equation \ref{eq:ChangeKinCueEight} must equal the energy gained by the eight ball, which is given by Equation \ref{eq:EightBallInitialKE}.

Equations \ref{eq:ChangeKinCueEight} and \ref{eq:EightBallInitialKE} agree within experimental uncertainty, so the collision between the cue ball and eight ball is elastic.  Since sound is heard during the collision, some of the energy is clearly lost.  Energy can also be converted to other forms such as heat and light.  The experimental uncertainties are large enough to accommodate this loss.

\section{Linear Momentum Conservation}

\begin{figure}[h!]
    \centering
       \includegraphics[width = 3.4 in, height = 2.5 in]{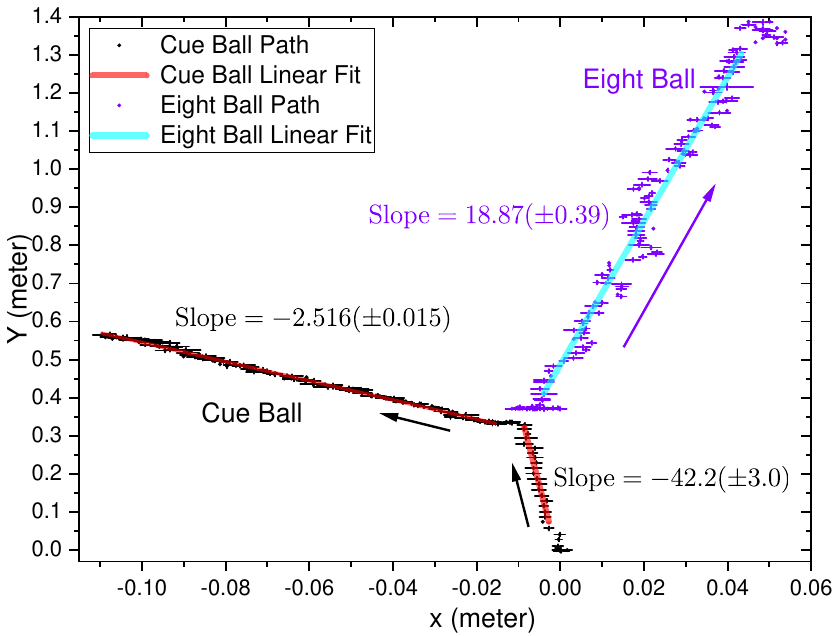}
    \caption{The digitized paths of the cue ball and eight ball (points) and a linear fit to the trajectories (lines).  Note that the $x$ and $y$ scales are different, so the angles are not to scale.  As a result, the uncertainties of the $x$ positions appear larger than for the $y$ positions, but they are about the same. }
    \label{plot:Paths}
\end{figure}
Finally, we test momentum conservation in the collision.  Since all of the action-reaction forces on the balls are equal and opposite, linear momentum should be conserved, though there are some subtleties that we will not discuss here.\cite{knight16.01,cross16.01}  Figure \ref{plot:Paths} shows the trajectories of the cue ball and the eight ball.  The points with error bars are the values obtained using Origin's Video Extractor and the lines are linear fits to the data.  Shown are the slopes associated with the three linear fits of the cue ball before the collision, the cue ball after the collision and the eight ball after the collision.  Note that the larger variations in the positions of the eight ball are due to the difficulty in seeing the center of a black ball on a dark background.  The swings in positions are on the order of a centimeter.  Even so, the uncertainty in the slope from the linear fit is about 2\%.

Using the speeds of the balls given in Tables \ref{tab:QuadParamCue} and \ref{tab:QuadParamEight}, we can calculate the components of the velocity using the speed and the angles determined from the slopes.  Since the slope $s$ is the tangent of the angle the velocity vector makes with the $x$-axis, we have
\begin{align}\label{eq:slopeTangent}
s^2 = \tan^2 \theta = \frac {\sin^2 \theta} {1 - \sin^2 \theta} .
\end{align}
Solving for $\sin \theta$, we get
\begin{align}\label{eq:slopeSine}
\sin \theta = \frac {s} {\sqrt{1 + s^2}}
\end{align}
and solving for $\cos \theta$ yields
\begin{align}\label{eq:slopeCosine}
\cos \theta = \sqrt{1 - \sin^2 \theta} = \frac {1} {\sqrt{1 + s^2}} .
\end{align}

\begin{table}\footnotesize
        \centering
       \caption{Velocities of the balls before (i) and after (f) the collision determined from the slopes and the speeds.\label{tab:velocity}}
        \begin{tabular}{c c c c c}
       \hline
         & Speed $v \, \mbox{(cm/s)}$ & Slope $s$ & $v_x\, \mbox{(cm/s)}$ & $v_y \, \mbox{(cm/s)}$ \\
      \hline \hline
      $\mbox{Cue}_i$ & $145.5(\pm 7.4)$ & $ -42.2(\pm 3.0)$ & $-3.45 (\pm 0.30)$ & $145 (\pm 13)$ \\
      $\mbox{Eight}_i$ & 0 & 0 & 0 & 0 \\
      $\mbox{Cue}_f$ & $-8.0(\pm 1.5)$ & $ 251.6 (\pm 1.5)$ & $ 3.0 (\pm 0.6)$ & $-7.4 (\pm 1.4)$ \\
      $\mbox{Eight}_f$ & $146(\pm 34)$ & $ 1887 (\pm 39)$ & $ -7.73 (\pm 0.18)$ & $146 (\pm 34)$ \\
      \hline
        \end{tabular}
    	\label{table:velocity}
    \end{table}
The components of the velocities are determined using Equations \ref{eq:slopeSine} and \ref{eq:slopeCosine}, yielding
\begin{align}\label{eq:Vx}
v_x = v \cos \theta = \frac {v} {\sqrt{1 + s^2}} \hspace{1em}  \mbox{ and } \hspace{1em} v_y = v \sin \theta =  \frac {vs} {\sqrt{1 + s^2}} .
\end{align}
Table \ref{tab:velocity} shows the slopes, the speeds and the velocities obtained from them for the cue ball and eight ball before and after the collision.  Appendix \ref{app:uncertainty} derives the prorogation of uncertainties in calculating the velocity components.

\begin{table}[h]
        \centering
       \caption{Total momentum before and after the collision.\label{tab:momentum}}
        \begin{tabular}{c c c}
       \hline
          & $p_x \, (\mbox{kg} \cdot \mbox{m/s})$ & $p_y \, (\mbox{kg} \cdot \mbox{m/s})$ \\
          \hline \hline
         Before & $0.00586 (\pm 0.00051)$ & $0.247 (\pm 0.022)$ \\
         After & $-0.0073 (\pm 0.0124)$ & $0.181 (\pm 0.055)$ \\
      \hline\\
        \end{tabular}
        \includegraphics{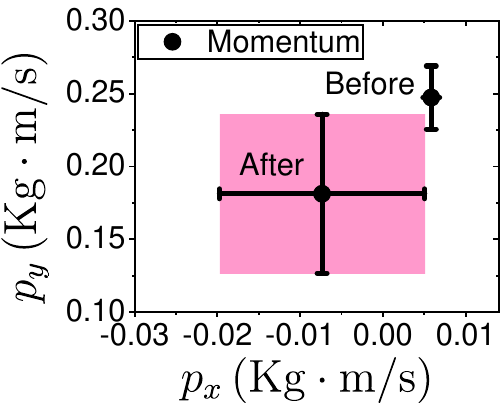}
    	\label{table:MomentumCinservation}
    \end{table}

Table \ref{tab:momentum} summarizes the momentum components before the collision, which resides in the cue ball, and the momentum components after the collision, which are the sum of the cue ball's and eight ball's momenta.  The figure below Table \ref{tab:momentum} plots these momenta.  The uncertainty range of the momentum after the collision is highlighted in pink, showing that linear momentum is conserved just at the edge of the uncertainty range.

There are several effects that we have not considered here that could account for some of the variations, such as inhomogeneities in the felt, an angular momentum component perpendicular to the table which can make the ball curve slightly while it is slipping, leading to an incorrect determination of the speed and angle.  We see some evidence of this in the eight ball's motion just after the collision in Figure \ref{plot:Paths}, where the initial trajectory appears to be slightly different than on the rest of the path.  A detailed analysis of such effects would be a distraction from the flow of this paper, so are not considered further.

\section{Designing a Lab}

The experiment presented here would be suitable for multiple laboratory reports that illustrate many concepts and physical principles.  As such, it could straddle multiple weeks.

To learn how to take data, digitize it and analyze the results, it would be appropriate for the students to start with the most simple experiment of a ball in free fall to determine the acceleration due to gravity and energy loss due to bouncing.  Even this simple experiment can lead to complications such as the role of parallax, errors introduced by the digitization process, getting the correct frame rate and the importance of careful controls when designing an experiment.  An additional bridge experiment might be a ball on a ramp that is covered with different materials so that the ball can be made to roll with and without slipping. Then students could move on to the kinds of experiments described here.

Since the data can be taken quickly in one session, the first lab report could focus on determining the velocities and accelerations from quadratic fits to various parts of the balls' trajectories.  Other considerations include determining how to get the boundaries and understanding why the derivative data are noisy.  The students could then be asked to make hypotheses about the causes of the accelerations and to point out peculiarities in the data.  The latter would be appropriate for group discussions or online chat rooms.

The second week could focus on applying Newton's laws to determine coefficients of friction, forces and torques.  The third week could apply the work-energy theorem for both translational and rotational motion and the forth week would study energy and momentum conservation of the collision.  There are also many possibilities for open-ended questions.  Students can go beyond the initial assignment by finding ways to determine the angular velocity, and using it to test the torque equation directly.

Students would learn about assigning uncertainties to measurements, doing fits and propagating error.  Also, free-body diagrams can be used to sure up a student's ability to visualize and solve a problem.

Alternatively, the experiment could be done on campus with pucks on an air table with the video provided to the student so that rolling is avoided.  After having mastered energy and momentum conservation in two dimensions, the student could proceed to the pool table experiments to add the various forms of friction, angular momentum and work-energy theorem.  Finally, markings added to the balls could be used to determine angular motion, or patches of other materials could be added to provided transitions to study friction and dissipation.

\section{Conclusions}

We have shown how the time dependence of the positions of billiard balls on a pool table, determined from a video taken with a mobile phone and digitized using the Video Extractor app in Origin reveals many principles of physics.  Most obviously, such experiments can be used to study energy and momentum conservation.  Indeed, the student can verify that energy and momentum are conserved in the collision.  In addition, we studied angular acceleration and deceleration due to the frictional torque from sliding and determined the coefficient of sliding and rolling friction.  All values fall in the range determined for typical billiard balls found by others.  Finally, we verified the Work-Energy Theorem.

Such experiments connect with students, who can see for themselves how physics applies to real-world systems that need not be limited to the highly-controlled physics laboratory.  Furthermore, simple video-based experiments make it possible for the student to work independently off campus and to be made more aware of the physics surrounding us all in our everyday lives.

A cell phone can be more generally applied to many other problems.  Past student projects included the analysis of Newton's cradle and the motion of a pendulum with high amplitude, masses sliding down inclined planes and curved surfaces, trajectories of rotating bodies with irregular shapes, the trajectory of a golf ball in the air and bouncing on the green, an analysis of Euler's Coin, a bouncing ball, the helium balloon pendulum in an accelerating vehicle and air resistance of various objects.  The variety of ideas is endless, giving students the opportunity to do hands-on experiments during these times when institutions of higher education are closed and considering hybrid labs in the future.

{\bf Acknowledgements:} We thank OriginLab Corporation\cite{origin2020} for providing free copies of Origin\cite{origin20.10} for use in our junior-level mechanics class, where this project started.  We are indebted to Yiming Chen from OriginLab for developing Video Extractor and making revisions based on our suggestions.  We also thank Trevor Foote and David Morin for their help.  Finally, we thank the reviewers for catching critical errors and making substantive suggestions that have resulted in a far better manuscript than the original.

\appendix

\section{Uncertainty Derivation}\label{app:uncertainty}

This appendix shows how we propogate uncertainties,\cite{bevin02.01} using the momentum conservation calculation for illustration.

We start by taking the differential of Equation \ref{eq:Vx}, yielding
\begin{align}\label{eq:dvx}
d v_x = \frac {dv} {\sqrt{1+s^2}} - \frac {sv \, ds} {(1+s^2)^{3/2}} .
\end{align}
Dividing Equation \ref{eq:dvx} by Equation \ref{eq:Vx} gives
\begin{align}\label{eq:dvx/Vx}
\frac {d v_x} {v_x} = \frac {dv} {v} - \frac {s\, ds} {1+s^2} .
\end{align}
Since the uncertainties add in quadrature, we get the uncertainty of the $x$-component of the velocity $\Delta v_x$ to be of the form
\begin{align}\label{eq:dvx/Vx}
\frac {\Delta v_x} {v_x} = \sqrt{\left(\frac {\Delta v} {v} \right)^2 + \left( \frac {s^2} {1+s^2} \right)^2  \left(\frac{\Delta s} {s} \right)^2},
\end{align}
where $\Delta v$ and $\Delta s$ are the uncertainties in the speed and slope.  Similarly, for $\Delta v_y$, we get
\begin{align}\label{eq:dvy/Vy}
\frac {\Delta v_y} {v_y} = \sqrt{\left(\frac {\Delta v} {v} \right)^2 + \left( \frac {s^2 - s + 1} {1+s^2} \right)^2  \left(\frac{\Delta s} {s} \right)^2}.
\end{align}

All uncertainties throughout this paper are computed using this procedure.


\end{document}